\documentstyle[aps,epsf]{revtex}
\tighten
\newcommand{\beq}{\begin{equation}}
\newcommand{\eeq}{\end{equation}}
\newcommand{\psibar}{\overline{\psi}}
\newcommand{\la}{\langle}
\newcommand{\ra}{\rangle}
\newcommand{\amp}[1]{\la #1 \ra}

\newcommand{\ba}{\begin{eqnarray}}
\newcommand{\ea}{\end{eqnarray}}
\newcommand{\dz}{\int \frac{d^{4}z}{(2\pi)^4}}
\newcommand{\dzp}{\frac{d^{4}z'}{(2\pi)^4}}

\newcommand{\slsh}[1]{\mbox{$\not\! #1$}}
\newcommand{\bm}[1]{\bbox{#1}}

\begin{document}

\textwidth 17 cm

\draft
\title{
\begin{flushright}
\begin{minipage}{3 cm}
\small
hep-ph/9710223\\
NIKHEF 97-038\\ 
VUTH 97-15
\end{minipage}
\end{flushright}
Single spin asymmetries from a gluonic background in the Drell-Yan process}
\author{D. Boer$^1$, P.J. Mulders$^{1,2}$ and O.V. Teryaev$^3$}
\address{\mbox{}\\
$^1$National Institute for Nuclear Physics and High--Energy
Physics (NIKHEF)\\	
P.O. Box 41882, NL-1009 DB Amsterdam, the Netherlands\\
\mbox{}\\
$^2$Department of Physics and Astronomy, Free University \\
De Boelelaan 1081, NL-1081 HV Amsterdam, the Netherlands\\
\mbox{}\\
$^3$Joint Institute for Nuclear Research, 141980 Dubna, Russia}

\maketitle
\begin{center} October 2, 1997\end{center}

\begin{abstract}
We discuss the effects of so-called gluonic poles in twist-three 
hadronic matrix elements, as first considered by Qiu and Sterman, in the 
Drell-Yan process. These effects 
cannot be distinguished from those of time-reversal odd distribution
functions, although time-reversal invariance is not broken by the 
presence of gluonic poles. Both gluonic poles and time-reversal odd 
distribution
functions can lead 
to the same single spin asymmetries.
We explicitly show the connection between gluonic poles and  
large distance gluon fields, identify the possible single spin asymmetries
in the Drell-Yan process and discuss the role of intrinsic transverse 
momentum of the partons.  
\end{abstract}


\section{Introduction}

In the usual description of the Drell-Yan process (DY) in terms of quark and
antiquark distribution functions time-reversal symmetry implies the absence of
single spin asymmetries 
at tree level, even including order $1/Q$ 
corrections \cite{Tangerman-Mulders-95a}.
Additional time-reversal odd (T-odd) 
distribution functions are 
present when  
the incoming hadrons cannot be treated as plane-wave states.
This may occur due to some factorization breaking mechanism \cite{Anselmino}. 
We will show that, even apart from such mechanisms, the 
contributions of T-odd distribution functions may effectively arise due 
to the presence of so-called gluonic poles attributed to asymptotic (large 
distance) gluon fields. The gluonic poles appearing in the 
twist-three hadronic matrix elements 
\cite{QS-91b,QS-92,Korotkiyan-T-94,E-Korotkiyan-T-95} together with 
imaginary phases of hard 
subprocesses effectively give rise to the same single spin asymmetries.
This is the
origin of the single spin asymmetry of Ref.\ \cite{Hammon-97}. 
Hence, the absence or
presence of single spin asymmetries in DY can be viewed as a reflection of
the absence or presence of gluonic poles.
The "effective" T-odd functions coming from  
gluonic poles do not constitute a violation of time-reversal invariance 
itself. 

The outline of the article is as follows. We will first discuss how DY is
described in terms of so-called correlation functions (section 2), 
which themselves are parametrized in terms of distribution functions 
(section 3). We focus especially on T-odd distribution functions, 
which show up in the
imaginary part of the equations of motion (e.o.m.), which relate quark 
correlation functions with and without an additional gluon. In section 4  
we will investigate the behavior of the quark-gluon correlation function 
in 
case it has a pole when the gluon has zero momentum. We will show that such
poles  
will effectively contribute to the imaginary part of the e.o.m.\ and 
hence, to T-odd distribution functions. 
The large distance nature of gluonic poles is elaborated upon, in
particular, the boundary conditions. In the final 
section we present the $Q_T$-averaged DY cross-section with emphasis on the 
contributions of 
the effective
T-odd distribution functions and the intrinsic transverse 
momentum. 

\section{The Drell-Yan process in terms of correlation functions}

We employ methods originating from Refs.\ 
\cite{Soper-77,Ralst-S-79,Coll-S-82,Politzer-80,EFP-83,Efremov-Teryaev-84,Jaffe-Ji-91,Jaffe-Ji-92} 
in order
to describe the soft (non-perturbative) parts of the scattering process in 
terms of correlation functions, which are (Fourier transforms of) hadronic
matrix elements of non-local operators. We restrict ourselves to tree-level, 
but 
include
$1/Q$ power corrections. The asymmetries under investigation are loosely
referred to as 'twist-three' asymmetries, since they are suppressed by a 
factor 
of $1/Q$, where the photon
momentum $q$ sets the scale $Q$, such that $Q^2=q^2$. We do not take $Z$
bosons into account, since the asymmetries are likely to be negligible at or
above the $Z$ threshold. 

For the Drell-Yan process up to order $1/Q$ the quark correlation functions to 
consider are \cite{Soper-77,Ralst-S-79,Coll-S-82,EFP-83}
\ba
&& \Phi_{ij}(P_1,S_1;p)=\dz e^{ip \cdot z} 
\amp{P_1,S_1|\psibar_j (0) \psi_i (z) | P_1,S_1},\\
&& \Phi^{\alpha}_{Aij}(P_1,S_1;p_1,p_2)=\dz \dzp e^{ip_1 \cdot z} 
e^{i(p_2-p_1) 
\cdot z'}
\amp{P_1,S_1|\psibar_j (0) g A^{\alpha}(z') \psi_i (z) 
| P_1,S_1}.
\ea
We have included a color identity and $g$ times $t^a$
from the hard into the soft parts $\Phi$ and $\Phi_A^\alpha$, respectively.
The inclusion of path-ordered exponentials, which are needed in order to 
render the correlation functions gauge invariant, is implicit.

The quark-quark correlation function $\Phi_{ij}(p)$ can be expanded in a 
number of
invariant amplitudes according to Dirac structure \cite{Ralst-S-79}. 
The available vectors are
the momentum and spin vectors $P_1, S_1$ of the incoming hadron (spin-1/2),
such that $P_1\cdot S_1=0$,  
and the quark momentum $p$. In the case of a hard scattering process the 
momentum of the struck quark is predominantly along the direction of the 
hadron momentum, which itself is chosen to be predominantly along a light-like
direction given by the vector $n_+$. Another light-like direction $n_-$ is 
chosen such that 
$n_+\cdot n_-=1$; both vectors are dimensionless. The second hadron is chosen 
to be predominantly in the $n_-$ direction, such that $P_1 \cdot P_2 = {\cal
O} (Q^2)$. 
We make the following Sudakov decompositions:
\begin{eqnarray}   
&& P_1^\mu \equiv \frac{ Q}{x_1 \sqrt{2}}\,n_+^\mu 
+ \frac{x_1 M_1^2}{ Q\sqrt{2}}\,n_-^\mu,\\
&& P_2^\mu \equiv \frac{x_2 M_2^2}{ Q\sqrt{2}}\,n_+^\mu
+ \frac{ Q}{x_2\sqrt{2}}\,n_-^\mu,\\
&&q^\mu \equiv \frac{ Q}{\sqrt{2}}\,n_+^\mu 
+ \frac{ Q}{\sqrt{2}}\,n_-^\mu 
+ q_T^\mu,
\end{eqnarray}
for $Q_T^2 \equiv - q_T^2 \ll Q^2$. 
We will often refer to the $\pm$ components of a momentum $p$, 
which are defined 
as $p^\pm=p\cdot n_\mp$.
Furthermore, we decompose the parton momenta $p, p_1$ and the spin vector 
$S_1$ of hadron-one as
\ba
&&p \equiv \frac{x Q}{x_1 \sqrt{2}}\,n_+
+ \frac{x_1 (p^2 + \bm{p}_T^2)}{x Q \sqrt{2}}\,n_- + p_T
\approx  x P_1 + p_T, \\[2 mm]
&&p_1 \equiv \frac{y Q}{x_1 \sqrt{2}}\,n_+
+ \frac{x_1 (p_1^2 + \bm{p}_{1T}^2)}{y Q \sqrt{2}}\,n_- + p_{1T} 
\approx  y P_1 + p_{1T}, \\[2 mm] 
&& S_1 \equiv \frac{\lambda_1 Q}{x_1 M_1 \sqrt{2}}\,n_+
-\frac{x_1 \lambda_1 M_1}{Q \sqrt{2}}\,n_- + S_{1T}
\approx  \frac{\lambda_1}{M_1} P_1 + S_{1T}^{}.
\ea
Also we note that up to order $1/Q$ only the
transverse components of $A^\alpha$ matter inside $\Phi_A^\alpha$. 
 
The Drell-Yan process consists of two soft parts and one of them is described
by the above quark correlation functions, whereas the other is defined by the
antiquark correlation functions, denoted by $\overline \Phi$ and 
$\overline \Phi{}_A^\alpha$. The correlation function $\overline \Phi$ depends
on the second hadron momentum and polarization, $P_2$ and $S_2$,  
and the antiquark momentum $k$ and is given by \cite{Ralst-S-79}:
\ba
\overline \Phi_{ij}(P_2, S_2;k) & = & 
\frac{1}{(2\pi)^4}\int d^4z\ e^{-i\,k\cdot z} 
\langle P_2,S_2 \vert \psi_i(z) \overline \psi_j(0) \vert P_2,S_2 \rangle.
\ea
The vectors in $\overline \Phi$ and 
$\overline \Phi{}_A^\alpha$ are also decomposed in $n_\pm$,
\ba
&&k \equiv \frac{\bar x Q}{x_2 \sqrt{2}}\,n_-
+ \frac{x_2 (k^2 + \bm{k}_T^2)}{\bar x Q \sqrt{2}}\,n_+ + k_T
\approx  \bar x P_2 + k_T, \\
&&S_2 \equiv \frac{\lambda_2 x_{2} Q}{M_2  \sqrt{2}}\,n_-
-\frac{\lambda_2 M_2}{x_{2} Q \sqrt{2}}\,n_+ + S_{2T}^{}
 \approx  \frac{\lambda_2}{M_2} P_2 + S_{2T}^{}.
\ea
The function $\overline \Phi_A^\alpha$ and the additional momentum $k_1$ are
analogously defined as for the quark case. 

At tree level four-momentum conservation fixes 
$x P_1^+=p^+=q^+=x_1 P_1^+$, i.e., $x=x_1$ and similarly $\bar x = x_2$, and 
allows up to 
$1/Q^2$ corrections for integration over $p^-$ and $k^+$. 
However, the transverse momentum integrations cannot be separated, unless one 
integrates
over the transverse momentum of the photon. In that case one arrives at 
correlation functions also integrated over their transverse momentum 
dependence, 
such that they only 
depend on the momentum fractions $x,y$ and $\bar x, \bar y$. These partly
integrated correlation functions $\Phi(x), \overline \Phi(\bar{x}), 
\Phi_A^\alpha(x,y)$ and $\overline \Phi{}_A^\alpha (\bar{x},\bar{y})$ are the 
quantities that are 
parametrized in terms of so-called distribution functions. 
For details see Ref.\ \cite{Tangerman-Mulders-95a}.

The five relevant diagrams lead to the following expression for the hadron
tensor integrated over the transverse photon momentum (up to order $1/Q$):
\begin{eqnarray}
\lefteqn{\int d^2 \bm{q}_T \, {\cal W}^{\mu\nu} = \frac{e^2}{3} \,\Biggl\{ 
\text{Tr}\left( \Phi (x) \gamma^\mu \overline \Phi 
(\bar{x}) \gamma^\nu \right)}\nonumber \\
&+& \int\, dy \, \text{Tr}\left( \Phi_A^\alpha (y,x) \gamma^\mu \overline \Phi 
(\bar{x}) \gamma_\alpha \frac{\slsh{n_+}}{Q\sqrt{2}} \, \frac{x-y}{x-y + i
\epsilon}\, \gamma^\nu \right) 
+ \int\, dy \, \text{Tr}\left( \Phi_A^\alpha (x,y)  \gamma^\mu
 \frac{\slsh{n_+}}{Q\sqrt{2}} \, \frac{x-y}{x-y + i\epsilon}\,
\gamma_\alpha \overline \Phi (\bar{x}) \gamma^\nu 
\right)
\nonumber \\[3mm]
&-& \int\, d\bar{y} \, \text{Tr} \left( \Phi (x) \gamma^\mu  
\overline \Phi_A^\alpha (\bar{y},\bar{x}) \gamma^\nu 
\frac{\slsh{n_-}}{Q\sqrt{2}} \, \frac{\bar{x} -\bar{y}}{\bar{x} -\bar{y}+ 
i \epsilon}\, \gamma_\alpha 
\right) 
- \int\, d\bar{y} \, \text{Tr}\left( \Phi(x) \gamma_\alpha
 \frac{\slsh{n_-}}{Q\sqrt{2}} \, \frac{\bar{x} -\bar{y}}{\bar{x} -\bar{y}+ 
i \epsilon}\, \gamma^\mu \overline \Phi_A^\alpha (\bar{x}, 
\bar{y}) \gamma^\nu 
\right) \Biggr\}.
\label{qtWmunu} 
\end{eqnarray}
We will explain the above expression. 
The factor 1/3 arises from the color averaging in the $q \bar q$ annihilation.
We have omitted flavor indices and summation; furthermore, there is a 
contribution from diagrams with reversed fermion flow, which is similar as the 
above expression but with $\mu \leftrightarrow \nu$ and $q \rightarrow
-q$ replacements.
In the expression the terms with $\slsh{n_\pm}$ arise from the fermion 
propagators in the hard part neglecting contributions that will appear 
suppressed by powers of $Q^2$,  
\begin{eqnarray}
\frac{\slsh{p_1}-\slsh{q}+m}{(p_1-q)^2-m^2+ i \epsilon} &\approx& 
-\frac{\slsh{n_+}}{Q\sqrt{2}}\, \frac{x-y}{x-y + i \epsilon},
\label{prop}
\\[3mm]
\frac{\slsh{q}-\slsh{k_1}+m}{(q-k_1)^2-m^2+ i \epsilon} &\approx& 
\frac{\slsh{n_-}}{Q\sqrt{2}} \, \frac{\bar{x} -\bar{y}}{\bar{x} -\bar{y}+ 
i \epsilon},
\end{eqnarray}
where the approximate signs hold true only when the propagators are embedded 
in the diagrams. From these expressions one observes that the case $x=y$,
i.e., the case of a zero-momentum gluon, corresponds to an on-shell quark 
propagator.  

Note that $\Phi(x),\overline \Phi(\bar{x}), \Phi_A^\alpha(x,y)$ and 
$\overline \Phi{}_A^\alpha (\bar{x},\bar{y})$ are now integrals involving 
only one 
light-cone direction, for instance, 
\begin{eqnarray}
\Phi_{ij} (x) & \equiv & \int \frac{d \lambda}{2\pi}e^{i\lambda x}\langle \,
P,S|\overline{\psi}_j (0) \psi_i(\lambda n_-)| P,S \, \rangle.
\ea

We observe that in the above expression one cannot simply replace 
$\Phi_A^\alpha(x,y)$ 
by $\Phi_D^\alpha(x,y)$, 
where
\beq
\Phi_{Dij}^{\alpha} (x,y) \equiv \int \frac{d \lambda}{2\pi} 
\frac{d \eta}{2\pi} e^{i\lambda x} e^{i\eta (y-x)} 
\amp{P,S|\psibar_j (0) iD_T^{\alpha}(\eta n_-) 
\psi_i(\lambda n_-)| P,S}
\eeq
and $iD^{\alpha}= i\partial^{\alpha} + g A^{\alpha}$. One must
takes into
account the difference proportional to 
$\int d^2 \bm{p}_T p_T^\alpha \Phi (x,p_T)$. This
difference is only zero, in case there are no transverse polarization vectors
present.
Similarly for the difference between 
$\overline{\Phi}{}_A^\alpha$ and $\overline{\Phi}{}_D^\alpha$. 

\section{Distribution functions}

For the correlation functions $\Phi$ and $\Phi^{\alpha}_{D}$ we need up to 
order $1/Q$ 
the following parametrizations 
in terms of distribution functions~\cite{Jaffe-Ji-91,Jaffe-Ji-92}:
\begin{eqnarray}
\Phi(x)&=&\frac{1}{2} \left[f_1(x) \mbox{$\not\! P_1\,$} + 
g_1(x)\,\lambda_1\gamma_{5}\mbox{$\not\! P_1\,$}
+ h_1(x)\,\gamma_{5}\mbox{$\not\! S$}_{1T}\mbox{$\not\! P_1\,$}
\right] \nonumber\\[3 mm]
&+& \frac{M_1}{2} 
\left[ e(x)\bm{1} + g_T(x) \gamma_5 \mbox{$\not\! S$}_{1T}
+ h_L(x) \frac{\lambda_1}{2} \gamma_5 \left[\slsh{n_+}, \slsh{n_-} \right]
\right],\label{paramPhi}\\[3 mm]
\Phi^{\alpha}_{D}(x,y)&=& 
\frac{M_1}{2} \bigg[ G_D(x,y)\,i\epsilon_T^{\alpha\beta} 
S_{1T \, \beta} \mbox{$\not\! P_1\,$} + \tilde{G}_D(x,y)\,
S_{1T}^\alpha \gamma_{5} \mbox{$\not\! P_1\,$}  \nonumber \\[3 mm]
&+& H_D(x,y) \lambda_1\gamma_{5} \gamma_T^\alpha \mbox{$\not\! P_1\,$}
+E_D(x,y)  \gamma_T^\alpha \mbox{$\not\! P_1\,$} \bigg],
\label{paramPhiD}
\ea
where $\epsilon_T^{\mu\nu}=\epsilon^{\alpha \beta\mu\nu} n_{+\alpha}
n_{-\beta}$.
We make a similar expansion for $\Phi_A^\alpha(x,y)$ with the
functions $G_D, \ldots$ replaced by $G_A, \ldots$, while the rest stays the
same. 

The parametrization of $\Phi(x)$ is consistent with requirements imposed 
on $\Phi$
following from hermiticity, parity and time-reversal invariance, 
\begin{eqnarray}
\Phi^\dagger (P_1,S_1;p) = \gamma_0 \,\Phi(P_1,S_1;p)\,\gamma_0 
& & \quad \mbox{[Hermiticity]} \\
\Phi(P_1,S_1;p) = \gamma_0 \,\Phi(\bar P_1,-\bar S_1;\bar p)\,
\gamma_0 & & \quad \mbox{[Parity]} \\
\Phi^\ast(P_1,S_1;p) = \gamma_5 C \,\Phi(\bar P_1,\bar S_1;\bar p)\, 
C^\dagger \gamma_5 & & \quad \mbox{[Time\ reversal]} 
\label{Tinvariance}
\end{eqnarray}
where 
$\bar p$ = $(p^0,-\bm{p})$, etc. 
For the one-argument functions in Eq.\ (\ref{paramPhi}) it follows from 
hermiticity 
that they are real. Note 
that for the validity of Eq.\
(\ref{Tinvariance}) it is essential that the incoming hadron is a plane wave
state. For $\Phi_D^\alpha$ and similarly for $\Phi_A^\alpha$  
hermiticity, parity and time-reversal invariance yield the following 
relations:
\begin{eqnarray}
\left[\Phi_D^\alpha(P_1,S_1;p_1,p_2)\right]^\dagger = 
\gamma_0 \,\Phi_D^\alpha(P_1,S_1;p_2,p_1)\,\gamma_0
& &\quad \mbox{[Hermiticity]} \\
\Phi_D^\alpha(P_1,S_1;p_1,p_2) = \gamma_0 
\,\Phi_{D\alpha}(\bar P_1,-\bar S_1;\bar p_1, \bar p_2 )\,
\gamma_0 & & \quad \mbox{[Parity]} \\
\left[\Phi_D^\alpha(P_1,S_1;p_1,p_2)\right]^\ast = 
\gamma_5 C \,\Phi_{D\alpha}(\bar P_1,\bar S_1;\bar p_1, \bar p_2  )\, 
C^\dagger \gamma_5 & &\quad \mbox{[Time\ reversal]} 
\label{TinvA}
\end{eqnarray}
Hermiticity then gives for the two-argument functions in Eq.\ 
(\ref{paramPhiD}) the following constraints:
\ba
G_D(x,y)& =&  -G_D^\ast (y,x),\\
\tilde{G}_D(x,y) &=& \tilde{G}_D^\ast (y,x),\\
H_D(x,y)& =&  H_D^\ast (y,x),\\
E_D(x,y)& =&  -E_D^\ast (y,x).
\ea
Hence, the real and imaginary parts of these two-argument functions have
definite symmetry properties under the interchange of the two arguments. 
If we would impose time-reversal invariance all four functions must be real
and $\tilde{G}_D$ and $H_D$ are then symmetric and 
$G_D$ and $E_D$ are antisymmetric
under interchange of the two arguments, such that at $x=y$ only $\tilde{G}_D$ 
and $H_D$ survive.

In the remainder of this section we do not impose time-reversal invariance 
and hence allow for 
imaginary parts of these functions. In addition, the following 
(T-odd) one-argument
distribution functions then appear:
\begin{eqnarray}
\left. \Phi(x) \right|_{\text{T-odd}}&=&\frac{M_1}{2} \left[ f_T(x)
\epsilon_T^{\mu\nu} S_{1T\mu}\gamma_{T\nu} - e_L(x) \lambda_1 i\gamma_5
+ h(x) \frac{i}{2} \left[\slsh{n_+},\slsh{n_-} \right]
\right]. 
\ea 
Also we parametrize:
\ba
\Phi_\partial^\alpha(x) \equiv \int d^2 \bm{p}_T p_T^\alpha 
\Phi (x,p_T) & = &
-\frac{M_1}{2} 
\bigg[ i f_{1T}^{\perp (1)} (x)\,i\epsilon_T^{\alpha\beta} 
S_{1T \, \beta} \mbox{$\not\! P_1\,$} - g_{1T}^{(1)} (x)\,
S_{1T}^\alpha \gamma_{5} \mbox{$\not\! P_1\,$}  \nonumber \\[3 mm]
&+& h_{1L}^{\perp (1)}(x) \lambda_1\gamma_{5} \gamma_T^\alpha \mbox{$\not\! 
P_1\,$}
+i h_{1}^{\perp (1)}(x)  \gamma_T^\alpha \mbox{$\not\! P_1\,$} \bigg].
\label{Phipartial}
\ea
The superscript $(1)$ stands for the first $\bm{k}_T^2$-moment of
$\bm{k}_T$-dependent distribution functions $f(x,\bm{k}_T^2)$,
\beq
f^{(1)}(x)= \int d^2 \bm{k}_{T} \, \left(\frac{\bm{k}_{T}^2}{2 M^2}\right) 
f(x,\bm{k}_{T}^2).
\eeq
This particular parametrization Eq.\ (\ref{Phipartial}) is written in a form
similar to Eq.\ 
(\ref{paramPhiD}), while using the $\bm{k}_T$-dependent functions of Ref.\
\cite{Mulders-Tangerman-96}. Note that $f_{1T}^{\perp (1)} (x)$ and
$h_{1}^{\perp (1)}(x)$ are T-odd. 

We observe (since $iD^{\alpha}= i\partial^{\alpha} + g A^{\alpha}$):
\ba
&& \int dy \, \left[ G_D(x,y) + G_D (y,x) \right]= 
\int dy \, \left[ G_A (x,y) + G_A (y,x) \right] - 2 i f_{1T}^{\perp (1)} (x),\\
&& \int dy \, \left[ \tilde{G}_D(x,y) + \tilde{G}_D (y,x) \right]= 
\int dy \, \left[ \tilde{G}_A (x,y) + \tilde{G}_A (y,x) \right] 
+ 2  g_{1T}^{(1)} (x),\\
&& \int dy \, \left[ H_D(x,y) + H_D (y,x) \right]= 
\int dy \, \left[ H_A (x,y) + H_A (y,x) \right] - 2 h_{1L}^{\perp (1)} (x),\\
&& \int dy \, \left[ E_D(x,y) + E_D (y,x) \right]= 
\int dy \, \left[ E_A (x,y) + E_A (y,x) \right] - 2 i h_{1}^{\perp (1)} (x),
\ea
while for the 'differences' no $\bm{k}_T^2$-moments appear:
\ba
&& \int dy \, \left[ G_D(x,y) - G_D (y,x) \right]= 
\int dy \, \left[ G_A (x,y) - G_A (y,x) \right],
\ea
etc. 

The two-argument functions and the one-argument functions are related by 
the classical e.o.m., which hold inside hadronic matrix
elements \cite{Politzer-80}. 
Using the above parametrizations one has the following relations
\cite{Efremov-Teryaev-84,Jaffe-Ji-92}: 
\ba
& & \int dy \, 
\Bigl[G_D(x,y)-G_D(y,x)+\tilde{G}_D(x,y)+
\tilde{G}_D(y,x) \Bigr] =2x g_T(x) - 2\frac{m}{M}h_1(x),
\label{eom}\\
& & \int dy \, 
\Bigl[G_D(x,y)+G_D(y,x)+\tilde{G}_D(x,y)-
\tilde{G}_D(y,x) \Bigr] = 2i x f_T(x),
\label{Imeom1}\\
& & \int dy \,  \Bigl[H_D(x,y)+H_D(y,x) \Bigr] = 
x h_L(x) -\frac{m}{M} g_1(x),\\
& & \int dy \,  \Bigl[H_D(x,y)-H_D(y,x) \Bigr] =
-i x e_L (x),
\label{Imeom2}\\
& & \int dy \,  \Bigl[E_D(x,y)-E_D(y,x) \Bigr] = 
x e(x) -\frac{m}{M} f_1(x),\\ 
& & \int dy \,  \Bigl[E_D(x,y)+E_D(y,x) \Bigr] = 
i x h(x)
\label{Imeom3}.
\ea 
From this we see that the (T-odd) imaginary parts of the 
two-argument functions are related to the T-odd one-argument 
functions, as one expects.
So if time-reversal invariance is imposed, the imaginary parts of the e.o.m.\ 
Eqs.\ (\ref{Imeom1}), (\ref{Imeom2})
and (\ref{Imeom3}) become three trivial equalities. We like to point out that
if one integrates Eqs.\ (\ref{eom}) and (\ref{Imeom1}) over $x$, weighted with 
some test-function $\sigma(x)$, one arrives at the sum rules discussed in 
\cite{Efremov-Teryaev-84,Teryaev-97}.  

In order to observe the role of intrinsic transverse momentum, we will use
some specific combinations of 
distribution functions, indicated by a tilde on the function. The tilde
functions are the true interaction-dependent twist-three parts of subleading
functions, which often contain twist-two parts, (in analogy to $g_2$) called 
Wandzura-Wilczek parts \cite{Wandzura-Wilczek-77}. 
They are defined such that in the analogues of Eqs.\ (\ref{eom}) --
(\ref{Imeom3}) for $G_A$ etc.\ only tilde functions appear, 
\ba
& & \int dy \, 
\Bigl[{\rm Re} \, G_A(x,y) + {\rm Re} \, \tilde{G}_A(x,y) \Bigr] =
x g_T(x) - \frac{m}{M} h_1(x) - g_{1T}^{(1)}(x) \equiv x \tilde{g}_T(x),\\
& & \int dy \, 
\Bigl[{\rm Im} \, G_A(x,y) + {\rm Im} \, \tilde{G}_A(x,y) \Bigr] = 
x f_T(x) + f_{1T}^{\perp (1)}(x) \equiv x \tilde{f}_T(x),
\label{Imeom1A}\\
& & \int dy \,  \Bigl[ 2 \, {\rm Re} \, H_A(x,y) \Bigr] = 
x h_L(x) - \frac{m}{M}g_1(x)
+ 2 \,  h_{1L}^{\perp (1)}(x) \equiv x \tilde{h}_L(x),\\
& & \int dy \,  \Bigl[2 \, {\rm Im} \, H_A(x,y) \Bigr] =
-x e_L(x) \equiv - x \tilde{e}_L(x), \\
& & \int dy \,  \Bigl[2 \, {\rm Re} \, E_A(x,y) \Bigr] = 
x e(x) - \frac{m}{M}f_1(x) \equiv x \tilde{e}(x),\\ 
& & \int dy \,  \Bigl[2 \, {\rm Im} \, E_A(x,y) \Bigr] = 
x h(x) + 2 \, h_1^{\perp (1)}(x) \equiv x \tilde{h}(x).
\label{Imeom3A}
\ea 

\section{Gluonic poles and time-reversal odd behavior}

\noindent We are interested in the behavior of the quark-gluon correlation 
function $\Phi_A^\alpha$ 
in case $x=y$, when the gluon has
zero-momentum. For this purpose, we define ($\alpha$ is a transverse index)
\beq
\Phi_{Fij}^\alpha(x,y) \equiv \int \frac{d \lambda}{2\pi} 
\frac{d \eta}{2\pi} e^{i\lambda x} e^{i\eta (y-x)} 
\amp{P,S|\psibar_j (0) F^{+\alpha} (\eta n_-) 
\psi_i(\lambda n_-)| P,S}
\eeq
and $F^{\rho\sigma}(z)=\frac{i}{g} \left[ D^\rho(z), D^\sigma(z) \right]$. 
Defined as given above, the matrix element has the same hermiticity,
but the opposite time-reversal 
behavior as $\Phi_D^\alpha$ and $\Phi_A^\alpha$ and we will parametrize it 
identically
with help of functions called $G_F(x,y), \tilde{G}_F(x,y), H_F(x,y)$ and
$E_F(x,y)$, noting that time-reversal implies (in contrast to $\Phi_D^\alpha$ 
or $\Phi_A^\alpha$) that $G_F$ and $E_F$ are symmetric and thus may survive at
$x=y$.
In the gauge $A^+=0$ one has $F^{+\alpha}=\partial^+ A_T^\alpha$ and one 
finds by partial integration 
\beq
(x-y)\Phi_A^\alpha (x,y)= -i \Phi_F^\alpha(x,y).
\label{AvsF} 
\eeq 
If a specific Dirac projection of $\Phi_F^\alpha(x,x)$ is nonvanishing, then 
the corresponding projection of $\Phi_A^\alpha(x,x)$ has a pole, hence the 
name gluonic pole. An example  
is the function $T(x,S_T)=\pi \text{Tr} \left[\Phi_F^\alpha(x,x)\, 
\epsilon_{T \beta \alpha} 
S_T^\beta \slsh{n_-} \right]/P^+ 
= 2\pi i M S_T^2 G_F(x,x)$ discussed by Qiu and 
Sterman in 
Ref.\ \cite{QS-91b,QS-92}. 

In order to define Eq.\ (\ref{AvsF}) at the pole, one needs a
prescription, which is related to the choice of boundary conditions on 
$A^\alpha(\eta = \pm \infty)$ inside matrix elements. 
Possible inversions of $F^{+\alpha}$ =
$\partial^+A_T^\alpha$ are (only considering the 
dependence on the minus component):
\ba
A_T^\alpha(y^-) &=& A_T^\alpha(\infty) 
- \int_{-\infty}^{\infty} dz^-\ \theta(z^--y^-)\,F^{+\alpha}(z^-)
\nonumber \\[3 mm]
&=& A_T^\alpha(-\infty) 
+ \int_{-\infty}^{\infty} dz^-\ \theta(y^--z^-)\,F^{+\alpha}(z^-)
\nonumber \\[3 mm]
&=& \frac{A_T^\alpha(\infty) + A_T^\alpha(-\infty)}{2}
- \frac{1}{2}\int_{-\infty}^{\infty} dz^-\ \epsilon(z^--y^-)\,F^{+\alpha}(z^-).
\ea
One can use the representations for the $\theta$ and $\epsilon$ functions,
to obtain 
\ba
\Phi^{\alpha}_A (x,y) &=& \delta(x-y)
\,\Phi^\alpha_{A(\infty)}(x)
+ \frac{-i}{x-y+i\epsilon}\,\Phi_F^\alpha(x,y)
\nonumber \\[3 mm]
&=&  
\delta(x-y)
\,\Phi^\alpha_{A(-\infty)}(x)
+ \frac{-i}{x-y-i\epsilon}\,\Phi_F^\alpha(x,y)
\nonumber \\[3 mm]
&=& 
\delta(x-y) 
\,\frac{\Phi^\alpha_{A(\infty)}(x) + \Phi^\alpha_{A(-\infty)}(x)}{2}
+ {\rm P}\frac{-i}{x-y}\,\Phi_F^\alpha(x,y),
\label{AvsFplusb}
\ea
where 
\beq
\delta(x-y)\,\Phi^\alpha_{A(\pm \infty)\,ij}
(x)
\equiv \int \frac{d \lambda}{2\pi} 
\frac{d \eta}{2\pi} e^{i\lambda x} e^{i\eta (y-x)} 
\amp{P,S|\psibar_j (0) gA_T^\alpha(\eta = \pm \infty) 
\psi_i(\lambda n_-)| P,S}.
\eeq
So Eq.\ (\ref{AvsFplusb}) shows the importance of boundary conditions
in the inversion of Eq.\ (\ref{AvsF}), if matrix elements containing
$A^\alpha(\eta = \pm \infty)$ do not vanish. When such matrix elements
vanish (implicitly assumed in \cite{Tangerman-Mulders-95a})
the pole prescription does not matter. 
Also one obtains 
\beq
2 \pi \, \Phi_F^\alpha(x,x) = 
\left[\Phi^\alpha_{A(\infty)}(x) - \Phi^\alpha_{A(-\infty)}(x)\right],
\eeq
which shows the
relation between the zero-momentum quark-gluon correlation function and the
boundary conditions. 

The behavior of $\Phi^\alpha_{A(\pm \infty)}(x)$ under time-reversal is:
\beq
\Phi^{\alpha \ast}_{A(\pm \infty)}(x)= \gamma_5 C
\,\Phi_{A(\mp \infty) \alpha}(x) \, C^\dagger \gamma_5.
\eeq
This relation implies that time-reversal invariance only allows for symmetric
or antisymmetric boundary conditions. 

To study the effect of gluonic poles 
we will consider the (nonvanishing) antisymmetric boundary 
condition\footnote{The consistency of antisymmetric boundary 
conditions with Maxwell's equations has already been shown in 
\cite{Kogut-Soper-70}.}, $\Phi^\alpha_{A(\infty)}(x)= 
- \Phi^\alpha_{A(- \infty)}(x)$, which implies $\pi\, \Phi_F^\alpha(x,x) =
\Phi^\alpha_{A(\infty)}(x)$.
In the diagrammatic calculation resulting in Eq.\ (\ref{qtWmunu}) 
one always encounters the pole of the matrix element (in this case in the 
principal value prescription) multiplied with the propagator in the 
hard subprocess (having a causal prescription),
\beq
\Phi_A^{\alpha \, \text{eff}} (y,x) \equiv 
\frac{x-y}{x-y+i\epsilon}\,\Phi_A^\alpha(y,x)
= \frac{-i}{x-y+i\epsilon}\,\Phi_F^\alpha(y,x)
= \Phi_A^\alpha(y,x) - \pi\,\delta(x-y) \,\Phi_F^\alpha(y,x).
\eeq
The time-reversal constraint applied to $\Phi_A^\alpha(x,y)$ implies 
the analogue of Eq.\ (\ref{TinvA}), while   
$\Phi_F^\alpha(x,y)$ has the opposite behavior under time-reversal compared to
$\Phi_A^\alpha(x,y)$. Thus for $\Phi_A^{\alpha \, \text{eff}} (x,y)$ one 
does not have definite behavior under T-reversal symmetry. 
Specifically, the allowed T-even functions of 
$\Phi_F^\alpha(x,x)$, $G_F(x,x)$ and $E_F(x,x)$, can be identified with T-odd 
functions in the effective 
correlation function $\Phi_A^{\alpha \, \text{eff}} (x,y)$. 
This implies that $G_A^{\text{eff}}(x,y)$ and $E_A^{\text{eff}}(x,y)$ will 
have an imaginary part and this gives rise to two 
"effective" time-reversal-odd distribution functions 
$\tilde{f}_T^{\text{eff}}(x)$ and $\tilde{h}^{\text{eff}}(x)$ via  
the (imaginary part of the) e.o.m.  
Since by identification
\ba
i \pi \, G_F(x,x)& =&  \int dy \,  
\text{Im} \, G_A^{\text{eff}}(y,x),\\
i \pi \, E_F(x,x) &= & \int dy \, 
\text{Im} \, E_A^{\text{eff}}(y,x),
\ea
it follows from Eqs.\ (\ref{Imeom1A}) and (\ref{Imeom3A}) that  
\ba
x \tilde{f}_T^{\text{eff}}(x) & =& i \pi G_F(x,x) = 
\frac{1}{2 M S_T^2} T(x,S_T), \\
x \tilde{h}^{\text{eff}}(x) 
&= &2 i \pi E_F(x,x) = \frac{- i \pi}{2MP^+}
\text{Tr}\left[ \Phi_F^\alpha (x,x)\, \gamma_{T \alpha} \slsh{n_-}\right].
\label{htildeproj}
\ea 
The function
$\tilde{e}_L^{\text{eff}}$ receives no gluonic pole contribution, since 
time-reversal symmetry requires $H_F(x,x)=0$. 

Of course, the mechanism for generating finite projections of 
$\Phi_F^\rho(x,x)$ remains unknown. We just can conclude that if
there is indeed a non-zero gluonic pole (in the case of 
non-zero antisymmetric boundary conditions), then at twist-three there are two
non-zero ``effective'' T-odd distribution functions, 
namely $\tilde{f}_T$ and $\tilde{h}$. 
The first one generates the single spin twist-three asymmetry found by Hammon\ 
{\em et al.} \cite{Hammon-97}, 
in their notation it is proportional to $T(x,x)$. The second one leads to a
new asymmetry (see next section). 
Summarizing, we find for the parametrization of
$\Phi^\alpha_{A(\infty)}(x)$, 
\ba
\Phi^\alpha_{A(\infty)}(x) &=& 
- \frac{i x M}{2}\bigg[ \tilde{f}_T^{\text{eff}}(x) 
\,i\epsilon_T^{\alpha\beta} 
S_{T \, \beta} \mbox{$\not\! P \,$} + 
\frac{1}{2} \tilde{h}^{\text{eff}}(x)
\gamma_T^\alpha \mbox{$\not\! P \,$} \bigg],
\label{paramPhiAinf}
\ea
which is constrained by time-reversal symmetry but behaves 
exactly
opposite to for instance $\Phi_\partial^\alpha (x)$, hence in their
parametrizations the meaning of time-reversal even or odd functions are
opposite also. 

The case of nonvanishing symmetric boundary conditions is less 
interesting, since $\Phi_F^\alpha(x,x)=0$, but it is allowed. 
The delta-function singularity in this 
case will contribute to the functions 
$\tilde{G}_A(x,x)$ and $H_A(x,x)$ and hence, to T-even tilde functions. This 
would only affect the magnitude of (time-reversal even) double spin 
asymmetries. 

The antisymmetric nonvanishing boundary condition for 
$\Phi^\alpha_{A(\pm\infty)}(x)$ might arise from a linear A-field, 
giving a constant field strength (cf.\ e.g.\ \cite{Schaefer-93,Ehrnsperger}). 
One might also think of an instanton background field. In both cases one 
should interpret infinity to mean 'outside the proton radius'. 
Also, the constant field strength should be understood as an average value
of the gluonic chromomagnetic field, which is non-zero due to a correlation
with the direction of the proton spin.
The large distance origin of the asymmetries arising from such a gluonic 
pole is apparent. 

We like to point out that so-called fermionic poles play a
role in off-forward scattering, such as prompt photon production 
\cite{ET-85,QS-91b,QS-92,Korotkiyan-T-94,E-Korotkiyan-T-95}, but not in
DY to this order. 

The fragmentation function that is the analogue of the distribution 
function $f_T$ (called $D_T$), shows up in 
a single spin asymmetry in hadron production in $e^+e^-$ annihilation 
\cite{Lu-95,Boer-97}, allowed because final state interactions lead 
to T-odd fragmentation functions. 
In Ref.~\cite{Lu-Li-Hu} both gluonic poles and final state interactions 
are considered, but without taking into account boundary terms in the matrix
elements. This result is in fact an example of the effective relation we 
have shown (see also \cite{Teryaev-97}). 

\section{The Drell-Yan cross-section in terms of distribution functions}

We will now discuss the Drell-Yan cross-section in case one integrates over
transverse photon momentum. So one uses the above parametrizations of the
correlation functions in the
expression for the integrated hadron tensor as given in Eq.\ (\ref{qtWmunu}), 
which after contraction with the 
lepton tensor yields the cross-section. The parametrizations in terms of
distribution functions are defined 
with help of the vectors $n_+, n_-$ and several transverse vectors. However, 
angles we are going to discuss with respect to another set of
vectors. Depending on the choice of this set, we find different combinations 
of functions with and without a tilde. Needless to say that the cross-section 
itself is an observable and does not depend on the choice of vectors, even 
though its appearance changes.  

We choose the following sets of normalized vectors:
\ba
& & \hat t \equiv q/Q,\\
& & \hat z \equiv  (1-c) \frac{2x_1}{Q} 
\tilde{P_1}- c \frac{2x_2}{Q} \tilde{P_2},\\
& & \hat x \equiv q_T/Q_T = (q-x_1\, P_1 -x_2\, P_2)/Q_T,
\ea
characterized by a parameter $c$ and where $\tilde{P_i} \equiv P_i-q/(2 x_i)$,
such that:
\begin{eqnarray}
n_+^\mu & = & 
\frac{1}{\sqrt{2}} \left[ \hat t^\mu + \hat z^\mu
-2c\,\frac{Q_T^{}}{Q} \hat x \right],
\label{transverse1} \\
n_-^\mu & = & 
\frac{1}{\sqrt{2}} \left[ \hat t^\mu - \hat z^\mu
- 2(1-c)\,\frac{Q_T^{}}{Q}\,\hat x^\mu \right]. 
\label{transverse2}
\end{eqnarray}
So the parameter $c$ basically distributes the transverse momentum between 
$P_1$ and $P_2$ in different ways (Fig.\ \ref{DYkin}). 
If $c=0$ ($c=1$), then $P_1$ ($P_2$) 
has no transverse component. The symmetric case $c=1/2$ is the one used in 
Ref.\ \cite{Meng-92}. 
\begin{figure}[htb]
\begin{center}
\leavevmode \epsfxsize=10cm \epsfbox{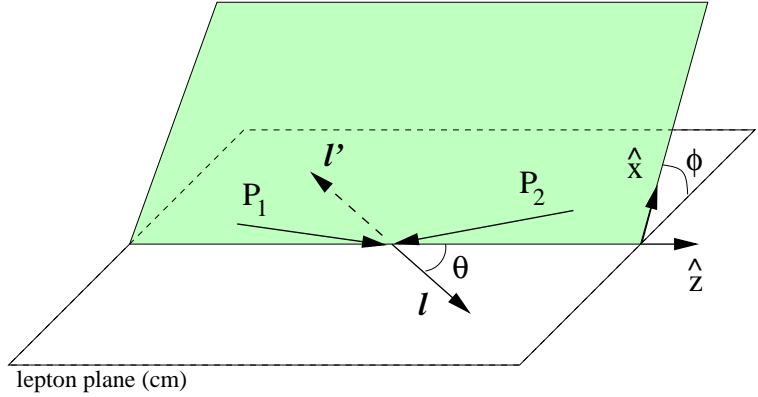}
\vspace{0.2 cm}
\caption{\label{DYkin} Kinematics of the Drell-Yan process in
the lepton center of mass frame, for a particular value of $c$.}
\end{center}
\end{figure}

In this way we arrive at the 
following expression for the Drell-Yan cross-section in case of
unpolarized leptons:
\begin{eqnarray} 
\lefteqn{\frac{d\sigma(h_1 h_2 \to l \bar{l} X)}{d\Omega dx_1 dx_2}=
\frac{\alpha^2}{3 Q^2}\;\sum_{a,\bar a}e_a^2\;\Bigg\{ 
       A(y) \left(
            f_1\overline f_1
          - {\lambda_1}{\lambda_2}g_1\overline g_1
         \right)}
\nonumber\\ &&
     {}+ B(y)\;|\bm S_{1T}^{}|\;|\bm 
S_{2T}^{}|\;\cos(\phi_{S_1}+\phi_{S_2})\;\left(
            h_1\overline h_1
         \right)  
\nonumber\\ &&
     {}+C(y)\;|\bm S_{1T}^{}|\;\sin(\phi_{S_1})
\;\bigg(
  \frac{2M_1}{Q}\, x_1 \, \left( (1-c)f_T + c \tilde f_T \right) \overline f_1
+ \frac{2M_2}{Q}\, x_2\, h_1 \left( c\overline h + (1-c)\tilde{\overline h} 
\right)
         \bigg)
\nonumber\\ &&
     {}+C(y)\;|\bm S_{2T}^{}|\;\sin(\phi_{S_2})
\;\bigg(
  \frac{2M_2}{Q}\, x_2\, f_1 \left( c\overline f_T + (1-c)\tilde{\overline f}_T
\right)
+ \frac{2M_1}{Q}\, x_1 \, \left(  (1-c) h + c\tilde h \right)\overline h_1
         \bigg)
\nonumber\\ &&
     {}+C(y)\;{\lambda_2}\;|\bm S_{1T}^{}|
\;\cos(\phi_{S_1})\;\bigg(
  \frac{2M_1}{Q}\, x_1 \, \big( (1-c) g_T + c\tilde g_T \big) \overline g_1
+ \frac{2M_2}{Q}\, x_2\, h_1 \left( c \overline h_L +(1-c) 
\tilde{\overline h}_L 
\right)
         \bigg)
\nonumber\\ &&
     {}-C(y)\;{\lambda_1}\;|\bm S_{2T}^{}|\;
\cos(\phi_{S_2})\;\bigg(
  \frac{2M_2}{Q}\, x_2\, g_1 \left( c \overline g_T + (1-c) 
\tilde{\overline g}_T
\right)
+ \frac{2M_1}{Q}\, x_1 \, \left( (1-c) h_L + c\tilde h_L \right) \overline h_1
         \bigg)
\Bigg\},
\end{eqnarray}
where
$d\Omega$ = $2dy\,d\phi^l$ and $\phi^l$ gives the 
orientation of $\hat l_\perp^\mu \equiv \left( g^{\mu \nu}-\hat t^{\{ 
\mu} \hat t^{\nu \} } + \hat z^{\{ 
\mu} \hat z^{\nu \} } \right) l_\nu$, the perpendicular part of the lepton
momentum $l$, and $y=l^-/q^-$.
In this result we encounter the following functions of $y$:
\ba
A(y) &=& \left(1-2y+2y^2\right)/2 
, \\
B(y) &=& y\,(1-y) 
,\\
C(y) &=& (1-2y)\,\sqrt{y\,(1-y)}.
\ea
Furthermore, $f_1 \overline f_1 = f_1^a (x_1)\overline{f}{}_1^a(x_2)$, etc.\ 
and where $a$ is the flavor index. 

For $c=1/2$ we find agreement with the results of \cite{Tangerman-Mulders-95a}
for the cross-section without T-odd distribution functions. Hence, we 
confirm the deviation of that result from the one found in \cite{Jaffe-Ji-92}. 

We observe single-transverse-spin asymmetries with two possible angular
dependences, namely $\sin(\phi_{S_1})$ and $\sin(\phi_{S_2})$. Each of them 
comes with two products of functions, in particular an unpolarized one ($f_1$
or $h$) times
a polarized one ($f_T$ or $h_1$). There is no choice of $c$ to eliminate the 
tilde functions from this expression, nor to only retain tilde
functions. This shows the non-trivial role of intrinsic transverse momentum of
the partons and one cannot discard it. This means that unlike in the case of 
DIS, one cannot take only $\Phi(x)$ and $\Phi_D^\alpha (x,y)$ as a basis
\cite{EFP-83}. 

If we assume that the presence of T-odd distribution functions is only
effective, arising due
to gluonic poles, and that $\Phi_{A(\infty)}^\alpha= 
\Phi_{D(\infty)}^\alpha$, then $\tilde{f}_T^{\text{eff}}= 
f_T^{\text{eff}}$ and $\tilde{h}^{\text{eff}}=h^{\text{eff}}$. This implies
the following single spin asymmetry (hadron-two unpolarized), given in the 
lepton center of mass frame:
\beq
A_T = \frac{2\sin(2\theta)\;\sin(\phi_{S_1})}{1 + \cos^2\theta} 
\frac{|\bm S_{1T}^{}|}{Q} \sum_{a}e_a^2
\;\bigg[2 M_1 \, x_1 \,f_T^a (x_1)
f_1^{\bar a}(x_2) + 2 M_2\, h_1^a(x_1) x_2 \, 
h^{\bar a}(x_2) 
\bigg]\Bigg/ \sum_{a}e_a^2 \; f_1^a (x_1) f_1^{\bar a} (x_2), 
\eeq
where we used that $y=(1 + \cos \theta)/2$ and $\theta$ is the angle of 
hadron-two with respect to the momentum of the 
outgoing leptons. The first term in the asymmetry (proportional to $f_T$) is 
the one discussed in
\cite{Hammon-97} (in their notation it is proportional to $T(x,x) q(y)$),
which will also be present in DIS ($f_1(x_2) = \delta(1-x_2)$).
The second term is 
the other, new single spin asymmetry arising in DY from a gluonic pole. It
is not proportional to $T(x,S_T)$, but to another projection of
$\Phi_F^\alpha$ in the point $x=y$, cf.\ Eq.\ 
(\ref{htildeproj}). 

\section{Conclusions}

We have shown how the effects of so-called gluonic poles in twist-three 
hadronic matrix 
elements, which were first discussed by Qiu and Sterman \cite{QS-91b,QS-92},
cannot be distinguished from that of T-odd distribution
functions. We investigated this 
for the Drell-Yan process, which is expressed in terms of products of
distribution functions. Even in the absence of T-odd distribution functions, 
imaginary phases arising from hard subprocesses together with gluonic poles
give rise to {\em effective\/} T-odd distribution functions. 
This leads to single spin asymmetries for the Drell-Yan process, such as the 
one found recently by Hammon {\em et al.\/} \cite{Hammon-97}. These
asymmetries therefore can have a 
different origin than the analogous
asymmetries in inclusive hadron production in $e^+e^-$ annihilation 
\cite{Lu-95,Boer-97}, which can also arise due to final state interactions, 
which are expected to be present always in contrast to initial state
interactions.
We have moreover shown that the presence of gluonic poles is in accordance 
with  
time-reversal invariance and requires a large distance gluonic field with
antisymmetric boundary conditions. 
Our analysis shows also the
role of intrinsic transverse momentum of the partons for the DY
cross-section at subleading order.  

\vspace{1cm}
We thank A. Sch\"afer for useful discussions. 
This work was in part supported by the Foundation for 
Fundamental 
Research on Matter
(FOM) and the National Organization for Scientific Research (NWO). 
It is also performed in the framework of Grant
96-02-17631 of the Russian Foundation for Fundamental Research
and Grant $N^o_-$ 93-1180 from INTAS.

\end{document}